# Effect of Moisture Absorption on Curing of Wind Blades during Repair

Sagar P. Shah, Michael N. Olaya, Evgenia Plaka, Joseph McDonald, Christopher J. Hansen, Marianna Maiarù*

*University of Massachusetts Lowell, Lowell, MA – 01854, USA.*

\_\_\_\_\_\_\_\_\_\_\_\_\_\_\_\_\_\_\_\_\_\_\_\_\_\_

\* Corresponding Author

# Abstract

Efficient structural repair of wind turbine blades is essential to limiting global warming and reducing the Levelized Cost of Energy (LCOE). Repairs carried out up-tower are sensitive to environmental conditions whose effect on the material properties during processing needs to be accounted for to accurately predict the repair outcome. This study investigates the effect of moisture content from environmental exposure on the cure kinetics of an infusion resin system used in wind turbine blade manufacturing and repair and provides an experimentally validated finite element tool for the analysis of cure cycle repairs as a function of repair geometry and moisture content. Moisture absorption tests on the two-part infusion system reported up to 12% moisture uptake by the curing agent under high temperature and relative humidity conditions. Differential scanning calorimetric measurements of resin in the presence of moisture revealed an accelerated cure behavior. Numerical predictions of a repair model agreed very well with the corresponding lab-scale repair and revealed a substantial temperature lag within the repair patch which resulted in thermal gradients and spatial distribution of the degree of cure. It was shown that the repair geometry and the accelerated-cure kinetics greatly influenced the temperature and cure distribution within the repair. The proposed approach can be used to reduce turbine downtime by minimizing the curing time.

1. **Introduction**

State-of-the-art wind turbine blades are composed of thermoset fiber-reinforced polymer composites (FRPC) that provide high mechanical performance and are lightweight due to their excellent specific mechanical properties [1], [2]. Over the last two decades, wind turbine size has progressively increased to harness wind energy in greater capacities and meet the global demand for increased production of sustainable energy [1], [3]. During their operational life of about 20-25 years, wind turbine blades are exposed to a variety of adverse environmental conditions, including lighting strikes, extreme winds, rains, thermal cycles, and foreign object impacts, all of which result in material degradation and structural damage [2]–[5]. Repairs are fundamental to restoring structural integrity and aerodynamic efficiency of the blades to ensure reliable wind turbine operations after the damage has occurred [6]. The operation and maintenance (O&M) cost accounts for a staggering 25% of the total levelized cost per kWh produced over the lifetime of a turbine [7], [8]. Repair procedures are highly dependent on the damaged area; when the repair size is manageable, the repair is performed up-tower in variable ambient conditions depending on the location, day, and time of the maintenance. Efficient and accurate computation techniques, able to predict the outcome of the repair and establish an optimized cure cycle that can ensure complete resin cure in minimum time, are needed to reduce the turbine downtime and increase the levelized cost of electricity (LCOE) for wind energy production.

After damage assessment and removal of the damaged substrate [1], [3], [6]–[8] the structural repair is performed by applying a scarf patch. The patch is made of a fiber-impregnated, viscous resin that needs to be cured within the parent laminate by applying temperature and pressure to restore the load-bearing capability of the material. Uniformity in the spatial distribution of the cure and temperature within the repair zone is critical to achieving high-quality repair. Aggressive cure cycles may lead to thermal gradients throughout the repair resulting in non-uniform curing of the constituent materials and overheating of the parent component, which may further introduce process-induced deformation, residual stress generation, and microcracking in the repair patch [1]. An in-depth understanding of the effect of processing parameters (including temperature ramp rates, hold temperature, and hold times) on the volumetric heat generation and distribution within the repair can avoid the aforementioned process-induced defects in the repair patch and aid in achieving ultimate structural integrity of the blade in a time- and cost-effective manner [9].

Computational modeling of structural repair can facilitate the prediction and assessment of composite processing for any repair geometry under given environmental conditions and provide repair-specific optimized procedures to achieve the highest quality repair [7], [8]. Various numerical approaches have reported the use of phenomenological kinetic models and three-dimensional heat transfer equations to predict cure evolution and temperature distribution during extensive blade repairs [6], [10]–[16]. Several studies have also employed constitutive relations to predict residual stress generation and distribution, assess the effect of defects and repair imperfections on strength and the lifetime of repaired structures, and optimize repair patch and scarfed joint configurations [7], [9], [17]–[23]. While such studies provide valuable insight into blade repair procedures, an aspect often ignored is the environmental conditions under which the repair is performed and its influence on the resin reactivity and mechanical behavior. The altered chemistry of the resin due to environmental exposure – relative humidity and temperature, for instance, may warrant a completely different set of processing conditions (compared to the standard operating procedures) for an effective repair. Therefore, it is critical to evaluate the influence of environmental exposure on resin behavior before a repair is planned [1], [24]–[27].

Several studies have reported the detrimental effect of moisture content on the polymeric resin's mechanical performance and reactivity [1], [3], [6], [24]–[28]. Pre-cure moisture absorption in resins and adhesives is known to accelerate the autocatalytic curing reaction, which in turn affects the volumetric heat generation [6], [25], [26]. Moisture uptake by the resin is highly dependent on the environmental conditions during repair and the exposure time, which makes the determination of an optimized cure cycle for the repair extremely challenging. Furthermore, degradation of the resin strength and stiffness due to phenomena such as plasticization, reduction in the glass transition temperature, and chemical swelling/degradation resulting from moisture uptake is widely reported [1], [24], [26]–[28]. Finally, moisture absorption by the parent laminate and the repair patch can introduce voids during the repair's curing phase, which can lead to delamination, rendering the repair ineffective [3]. Therefore, for a successful repair, the environmental and processing conditions and their influence on the structural integrity of the repair must be considered.

The objective of this study is to investigate the effect of moisture content from environmental exposure on the cure kinetics of an infusion resin system used in wind turbine blade manufacturing and repair and provide experimentally-validated computational tools for the analysis of cure cycle repairs. First, the effect of moisture on resin cure kinetics is assessed by carrying out a series of moisture absorption tests where the two-part resin system is exposed to several environmental conditions. Subsequently, the resin cure kinetics is characterized for various moisture content cases. Informed by the acquired cure kinetic data, a 3D finite element (FE) curing model is implemented within the commercial software Abaqus using user-written subroutines to virtually cure a repair patch and predict the degree of cure and temperature evolutions during the procedure. Numerical results from the simulations are experimentally validated in terms of temperature distribution and degree of cure evolution. Lastly, the numerical model is used to understand the influence of repair geometry and environmental conditions on the spatial distribution of temperature and degree of cure within the repair.

This work is a milestone towards developing a computationally-based integrated computational materials engineering (ICME) framework for optimizing repair procedures that will improve the repair quality and ensure the repaired blade's long-term structural integrity. The focus of this manuscript is the prediction of the degree of cure and temperature evolutions during repair with the understanding that temperature and cure gradients affect the mechanical properties evolution of the resin and therefore residual stress generation [12]. The prediction of residual stress generation and mechanical performance after repair are not the subjects of this study. Future work will embed the curing models derived in this study within an ICME framework that links material models, structural models, and experiments at multiple length sccassa to obtain digital twins of the blade [9], [21]–[23], [29]–[37].

This manuscript is organized as follows: the methodology to study the influence of moisture on cure kinetics for resin infusion and their corresponding curing models are presented in Section 2. Modeling results and their validation for two different scarfing slopes and different moisture content are detailed in Section 3. The main conclusions are summarized in Section 4.

## 2. Methodology

During in-field repairs, the resin mixture may get exposed to environmental moisture for extended periods, which may alter the curing kinetics. First, the moisture absorption was quantified according to O&M instructions. Then, the cure kinetics was characterized through Dynamic Scanning Calorimetry (DSC) for a widely used commercial epoxy system developed by Westlake Epoxy for wind energy applications: the infusion system EPIKOTE$^{TM}$ Resin MGS RIMR 135 with EPIKURE$^{TM}$ Curing Agent MGS RIMH 1366 (henceforth referred to as RIM R135-H1366). RIM R135-H1366 is a relatively low-viscosity system with an elevated curing temperature ideal for vacuum-assisted infusion processes. The system comprises bisphenol-A base epoxy resin with an amine-based curing agent. Unidirectional (UD) stitched E-glass fibers (areal weight of 955 g/m2), manufactured by Saertex and distributed by Fibre Glast Developments Corp., were used as reinforcements. The moisture absorption tests are detailed in Section 2.1. Section 2.2 describes the cure kinetic characterization. The curing model implementation is discussed in Section 2.3.

### 2.1 Moisture Absorption Characterization

The moisture absorbed by the base epoxy and the curing agent during the repair was characterized by separately exposing the two constituents to environmental conditions encountered during in-field repairs. 60 g of RIM R135 epoxy samples and 20 g of RIM H1366 curing agent samples were placed inside the environmental chamber in separate containers. The chamber was set to several combinations of temperature (5°C, 23°C, and 38°C) and relative humidity (RH of 50% and 90%). The initial specimen weight was recorded before the test, and subsequently, the specimens were weighed in five-minute intervals over a two-hour exposure period. This exposure window was chosen based on the analysis of wind turbine repair reports provided by Pattern Energy and TPI Composites. The relative percentage weight gain calculated for each data point in the sample reflected the moisture absorbed over the exposure period.

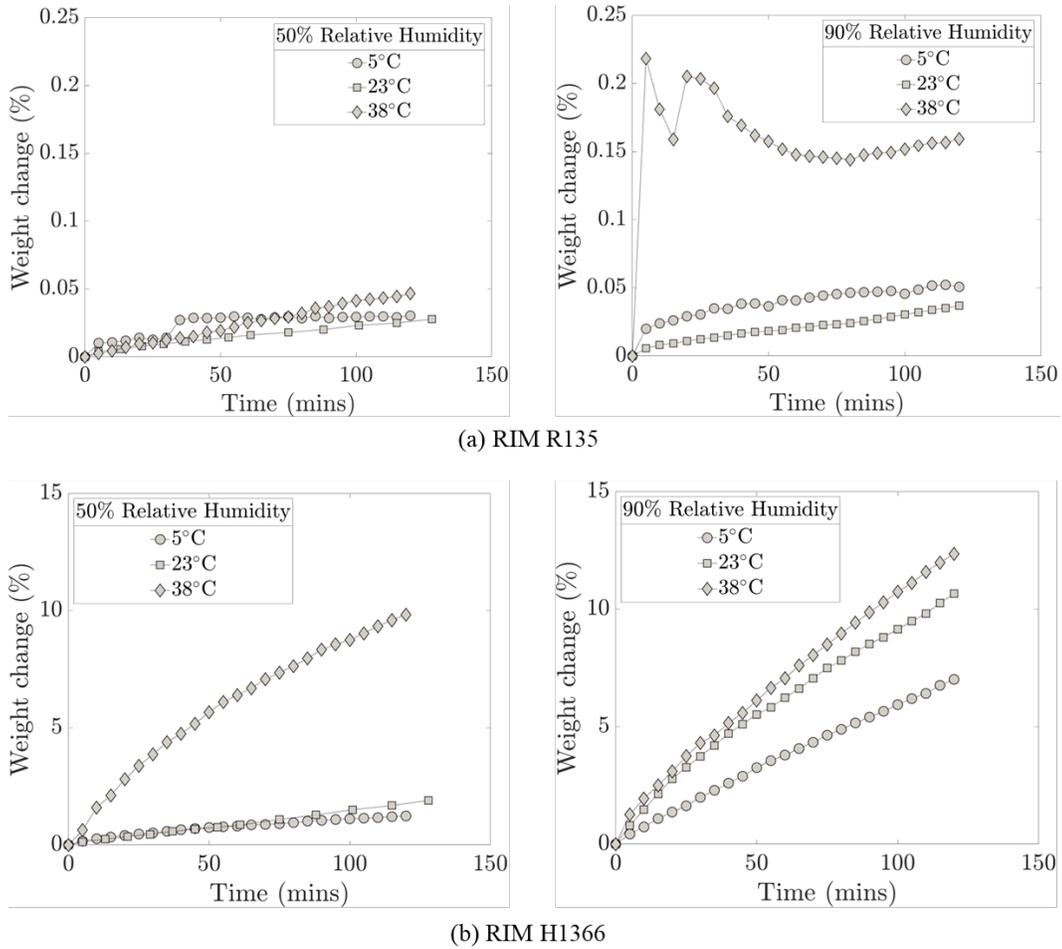

*Figure 1: Moisture absorption test results showing the percentage weight change in (a) base epoxy RIM R135 and (b) curing agent RIM H1366 when subjected to various temperature and relative humidity conditions.*

Figure 1 shows the moisture absorption test results for all six test conditions for both resin and curing agent. The RIM R135 resin Figure 1(a) revealed a minimal weight change of the conditioned RIM R135 epoxy resin, which suggested no moisture absorption. This result was expected due to the relatively high viscosity of the resin and the bisphenol-A chemistry, which prevented moisture absorption. On the contrary, a significant moisture absorption was displayed by the amine-based curing agent RIM H1366, as shown in Figure 1(b). This result was attributed to the low viscosity of the curing agent and the hydrophilic nature of the amine groups in the curing agent that allowed the absorption of the surface moisture deposition. A maximum moisture content of 12% relative to the initial weight was observed in the RIM H1366 when exposed to 38°C at 90% RH. Relatively lower moisture content was recorded for other test conditions, as summarized in Table 1. The cure kinetics of the mixed resin with the curing agent was then evaluated using DSC for the six case studies of 50% and 90% relative humidity, each at 5°C, 23°C, and 38°C.

*Table 1: Percentage of moisture absorption by base epoxy RIM R135 and curing agent RIM H1366 when subjected to various environmental conditions.*

| Temperature (°C) | Relative Humidity (%) | Moisture Absorption (%) RIM R135 | Moisture Absorption (%) RIM H1366 |
|---|---|---|---|
| 5 | 50 | 0.03 | 1.23 |
| 5 | 90 | 0.05 | 7.01 |
| 23 | 50 | 0.04 | 3.07 |
| 23 | 90 | 0.04 | 10.66 |
| 38 | 50 | 0.05 | 9.82 |
| 38 | 90 | 0.22 | 12.35 |

*2.2 Cure Kinetic Law*

The impact of moisture content on the resin cure kinetics was experimentally characterized using DSC (Discovery, TA Instruments). Dynamic scans were performed on small amounts of the uncured mixture (100:30 parts by weight) sealed in a hermetic aluminum pan weighing approximately 5 mg. The DSC specimens were first equilibrated at -10°C within the DSC chamber for two minutes to avoid premature curing. The chamber temperature was then increased from the equilibration temperature to 250°C at three different temperature rates of 5°C/min, 10°C/min, and 15°C/min. The exothermic heat of the reaction of the curing resin was measured as a function of time and temperature for each dynamic scan, as seen in Figure 2(a). The instantaneous heat of reaction, $dH/dT$, was obtained as the normalized heat flow at any given time, $t$. The partial heat of reaction, $H(t)$, up to time $t$ and the total heat of reaction, $H_T$, were computed by integrating the area under the normalized heat flow versus time plots up to time $t$ and the end of the reaction, respectively.

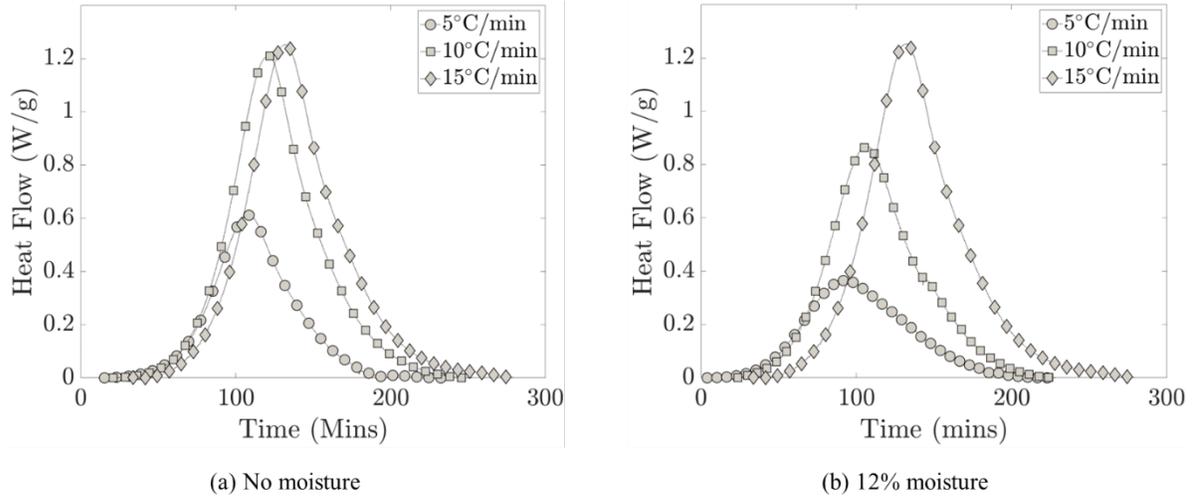

(a) No moisture

(b) 12% moisture

*Figure 2: Normalized heat flow versus time plots from dynamic DSC measurements taken at three different heating rates for (a) no moisture and (b) 12% moisture content case.*

The rate of curing, $d\phi/dt$, and the degree of cure, $\phi$, for each temperature ramp were then calculated based on $dH/dt$ at any time $t$ and $H_T$ given off by the curing thermoset as in Equations (1) and (2), respectively.

$$\frac{d\phi}{dt} = \frac{1}{H_T}\frac{dH}{dt} \quad (1)$$

$$\phi(t) = \frac{1}{H_T}\int_0^t \frac{dH}{dt}\,dt = \frac{H(t)}{H_T} \quad (2)$$

The rate and degree of cure obtained from Equations (1) and (2) were fitted to the autocatalytic Prout-Thompkins phenomenological kinetic model to determine the kinetic constants,

$$\frac{d\phi}{dt} = A\exp\left(-\frac{E_a}{RT^*}\right)[\phi^m(1-\phi)^n] \quad (3)$$

In Equation (3), four material-specific parameters, namely the activation energy $E_a$, pre-exponential factor $A$, and the dimensionless modeling parameters $m$ and $n$, were determined through a data fitting procedure from the experimental data. Here, $R$ is the universal gas constant and $T^*$ is the absolute temperature in Kelvins. An in-house data fitting tool was developed using MATLAB to perform nonlinear least-square analyses of the experimental data to determine the material-specific kinetic constants. The Kissinger method was used to provide initial estimates for $E_a$ and $A$ [6], [9] and the MATLAB "*lsqnonlin*" function was used for fitting. The best fits obtained for all temperature rates yielded the kinetic constants summarized in Table 2. The derived model predictions of the degree of cure as a function of time, $\phi(t)$, for the unconditioned specimen calculated using Equation (3) were compared to re-runs in DSC to determine any residual heat of reaction. A strong match was observed between the experiments and model predictions, which validated the kinetics model.

*Table 2: Cure kinetic modeling parameters for various moisture contents.*

| Moisture Content (%) | Total Heat of Reaction $H_T$ (J/g) | Kinetic Constants | | | |
|---|---|---|---|---|---|
| | | $E_a$ (kJ/mol) | $A$ (sec$^{-1}$) | $m$ (-) | $n$ (-) |
| 0 | 413.82 | 46.67 | 11000 | 0.3 | 1.5 |
| 2 | 392.62 | 38.44 | 750 | 0.4 | 1.4 |
| 7 | 361.28 | 55.4 | 450000 | 0.25 | 1.75 |
| 9 | 337.1 | 30.45 | 100 | 0.45 | 1.3 |
| 10 | 390.43 | 39.61 | 2250 | 0.25 | 1.35 |
| 12 | 364.75 | 29.3 | 60 | 0.38 | 1.2 |

The DSC tests and the data fitting procedure to determine the cure kinetics were repeated for conditioned specimens with varying moisture contents up to 12%. Deionized water was added to the resin mixture to condition the specimens in an amount corresponding to the values summarized in Table 1. Results for the maximum moisture content case of 12% are presented in Figure 2(b) and compared with those of the unconditioned specimens. The unconditioned specimens manifested a relatively higher exothermic peak spread over a narrower temperature range than the 12% moisture content, as summarized in Table 2. Furthermore, adding 12% moisture reduced the activation energy, $E_a$, and the pre-exponential factor, $A$, (see Table 2). Lower activation energy and pre-exponential factors indicated that the reaction required less energy input to begin autocatalyzing and a shorter time to approach exponential growth, beyond which the cure reaction slowed down. Figure 3 compares the degree of cure evolution of the unconditioned specimen as a function of time with the conditioned specimens with varying moisture contents for a prescribed cure cycle. Figure 3 shows that the degree of cure evolution is accelerated with increased moisture content. While the unconditioned specimen required 250 mins to achieve a full cure state ($\phi = 1$) with the prescribed temperature profile, the resin with 12% moisture content cured in 100 mins for the same cycle. These trends were consistent with the findings of Sharp et al. [26]. They reported an increase in molecular diffusion in uncured epoxy due to moisture, which led to an accelerated cure. However, after sufficient crosslinks had formed, the moisture in the epoxy formed pockets that hindered further molecular diffusion, significantly reducing the cure rate. The cure kinetics derived as a function of moisture content can be used to understand the effect of moisture on the curing of wind repair patches, as described in Section 3.

## 2.1 Process Modeling of Blade Repair

Process modeling of the wind blade repair is crucial to understand the effect of the cure cycle given the size and geometry of the repair, processing parameters, and environmental conditions, including temperature and moisture. The proposed virtual curing procedure is highlighted hereafter, including an in-depth description of the prescribed temperature profile and appropriate boundary conditions used in the simulations. The numerical model was generated within the FE software Abaqus supplemented by user-written subroutine UMATHT to simulate the curing process. Subsequent paragraphs describe the model geometry and the repair patch manufactured in the lab, detail the procedure to virtually recreate the patch model, including the laminate's stacking sequence and the material's thermal properties used for the repair, and present the experimental procedure performed in the laboratory to validate the FE model predictions. The results from the numerical and experimental study are presented in Section 3 in terms of the effect of scarfing slope and moisture content on the time required to fully cure the repair. Here, numerical results of curing the patch in the presence of moisture are compared against the same repair assuming no moisture intake in the resin.

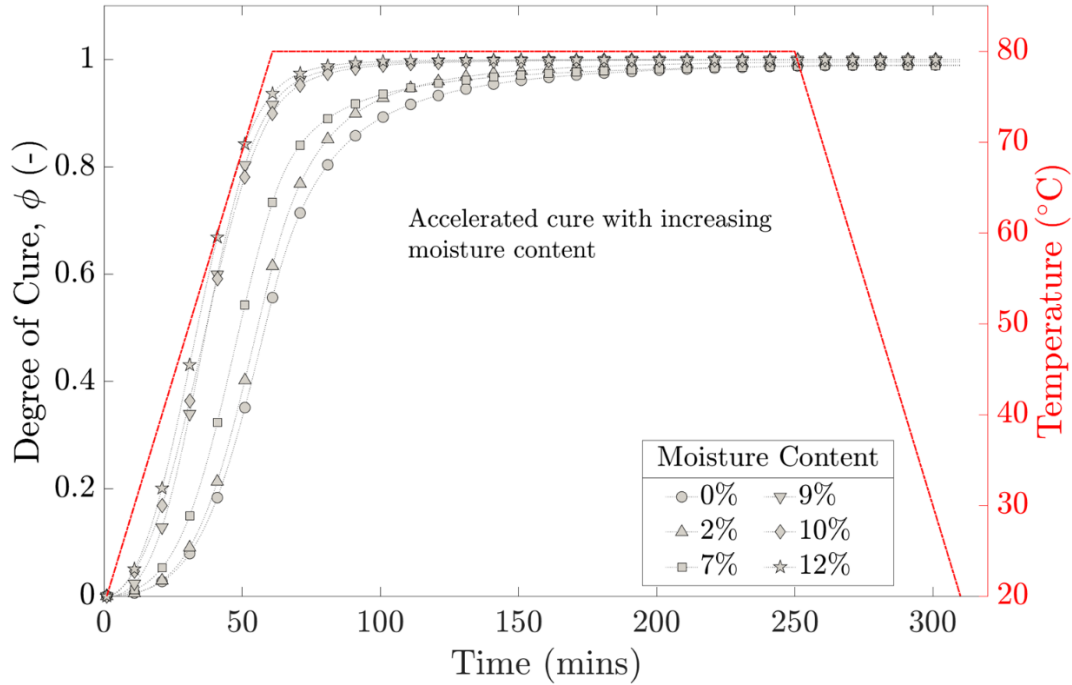

*Figure 3: Evolution of the degree of cure for the prescribed temperature profile as computed by the cure kinetic model for unconditioned (no moisture) and conditioned (various moisture content, 2% - 12%) specimens.*

A representative scarf repair patch geometry was modeled and virtually cured using an FE process modeling approach to evaluate the influence of the moisture and geometry of the patch on the curing time and degree of cure of the repair. A scarf geometry representative of a small structural up-tower repair was chosen for this analysis, as illustrated in Figure 4. The repair patch was manufactured in the lab and cured in the oven. Several thermocouples were embedded within the laminate through the repair patch thickness and along the scarf to monitor the temperature evolution, as illustrated in Figure 4. An additional thermocouple was also placed within the oven, external to the patch, to track the oven temperature throughout the repair process.

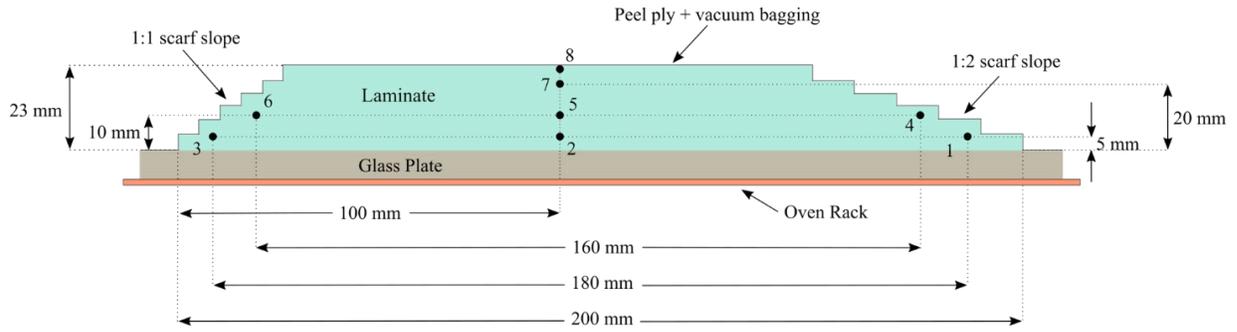

*Figure 4: Repair patch geometry for process modeling and lab-scale repair.*

The repair geometry featured a 200 mm base laminate (2 mm deep) made of E-glass fiber infused with RIM R135-H1366 epoxy resin, stacked up by ply-dropping plies asymmetrically to induce a different scarfing slope on each side. One side of the repair patch was manufactured with a scarfing ratio of 1:1, while the other was given a scarfing ratio of 2:1 to evaluate the influence of different geometrical scarfing ratios on the temperature and degree of cure distribution. The repair patch model was 23 mm in height. The thickness of the laminate was selected to simulate a repair that would exhibit an appreciable exotherm during curing and therefore measure a thermal gradient within the repair patch. The scarfing surfaces were not smoothened after manufacturing, as shown in Figure 4. The numerical model included a glass plate that supported the laminate, the oven shelf, the peel ply, and the vacuum bagging material required for infusion to closely replicate the lab-scale repair. The thermal properties used in the simulations are summarized in Table 3. The infusion was modeled by homogenizing the UD E-glass fibers with properties of the infused RIM R135-H1366 resin system, whose thermal properties were assumed to vary linearly as a function of the degree of cure, as shown in Equations (4) and (5).

$$c_p = c_p^0 + \phi * (c_p^1 - c_p^0) \tag{4}$$

$$k = k^0 + \phi * (k^1 - k^0) \tag{5}$$

where $c_p^0$ and $k^0$ are the specific heat capacity and thermal conductivity of the uncured resin, respectively, while $c_p^1$ and $k^1$ are the specific heat capacity and thermal conductivity for the fully cured resin, respectively. $\phi$ is the degree of cure. The composite material properties were homogenized, using the rule of mixtures [38], based on the constituent material properties summarized in Table 3 and the fiber volume fraction. The fiber volume fraction of the laminate was measured to be $v_f = 0.52$.

Table 3: Constituent material properties used for the process modeling of the repair.

| Material | Property | | |
|---|---|---|---|
| | Density, $\rho$ (g/cm³) | Thermal Conductivity $k$ ($\times 10^{-4}$ W/mm-K) | Specific Heat Capacity $c_p$ (J/g-K) |
| Borosilicate Glass Plate | 2.23 | 12 | 1.5 |
| Type 304 Stainless Steel | 7.93 | 163 | 1.4 |
| Patch Peel Ply Film | 1.14 | 2.5 | 1.75 |
| Vacuum Bag | 1.14 | 2 | 1.3 |
| RIM R135-H1366 ($\phi = 0$) | 1.20 | 1.59 | 1.35 |
| RIM R135-H1366 ($\phi = 1$) | 1.20 | 2.45 | 1.62 |
| E-glass Fiber | 2.55 | 12 | 1.2 |

The virtual curing analysis was carried out in Abaqus/STANDARD supplemented by a user-written subroutine, UMATHT. A thermal step was defined where the Fourier heat law of thermal conduction, with the addition of a heat generation term, $\dot{Q}$ was solved simultaneously with the Prout-Thompkins kinetic model to predict the heat released due to the progression of the cure reaction.

$$\rho^c c_P^c \frac{\mathrm{d}T}{\mathrm{d}t} = k^c \nabla^2 T + \dot{Q} \tag{6}$$

$$\dot{Q} = V^m \rho^m H_T^m \frac{\mathrm{d}\phi}{\mathrm{d}t} \tag{7}$$

Here, $\rho^c$, $c_P^c$ and $k^c$ are the composite density, specific heat capacity, and thermal conductivity of the curing patch, respectively determined based on the specified fiber volume fraction; $T$ is the temperature; and $V^m$, $\rho^m$ and $H_T^m$ are the resin volume fraction, density, and total enthalpy of the reaction, respectively. As depicted in Figure 5, thermal boundary conditions were prescribed to the model to mimic the curing process inside a convection oven. A temperature ramp from room temperature to 80°C at 1.5°C/min, isothermal hold at 80°C for 350 mins, and passive cooling to ambient conditions after removal from the oven was prescribed to the shelf during the virtual curing as per a thermocouple reading placed in the oven rack. Convective boundary conditions were prescribed to the faces of the model exposed to forced airflow inside the oven. The convective coefficients of the air in the oven were calculated on the different sides of the laminate and applied through convective film coefficients as shown in Figure 5 where $h_1 = 0.005$ W/mm²K, $h_2 = 0.01$ W/mm²K and $h_3 = 0.015$ W/mm²K, respectively. The model was meshed using 36753 DC3D8, eight-node linear heat transfer elements.

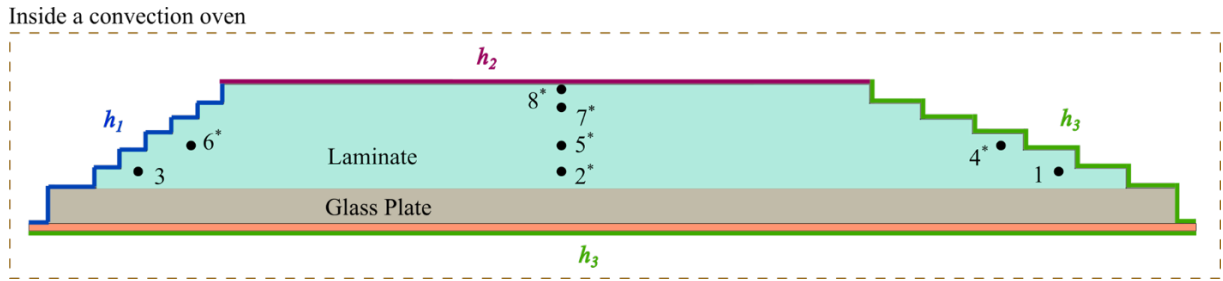

*Figure 5: Illustrations of the boundary conditions used for process modeling of the repair. Note that only thermocouples denoted with an asterisk (\*) provided readable data.*

## 3. Modeling Results and Experimental Validation

The repair patch was manufactured in the laboratory to validate the process modeling predictions. Thermocouples were embedded in the laminate to record temperature data throughout the patch and assess the spatial variation in the temperature due to the exothermic heat of the reaction and the resulting degree of cure, as shown in Figure 4. Temperature data during curing was compared to the numerical model predictions of the nodal temperature in four locations, as shown in Figure 5. Figure 6 compares the temperature evolution at specific nodes (4-7) from the virtual repair to the corresponding thermocouple data. Numerical model predictions agreed very well with the thermocouples read. Minor deviations in the magnitude and time to peak temperature were observed, which could be attributed to small imprecisions in defining the location of the thermocouple with respect to the computational nodes. The degree of cure evolution was computed from the temperature evolution data presented in Figure 6 per the Prout-Thompkins kinetic model in Equation (3), as shown in Figure 7.

### *3.1 Effect of Scarf Geometry on Blade Repair*

Two distinct scarfing slopes of 1:1 and 2:1 were analyzed to evaluate the influence of the scarf geometry on the temperature and cure distribution. Figure 8(a) presents an overlay of the temperature evolution at several nodes considered in this study. The plot clearly shows a spatial variation in the temperature within the repair patch. This spatial variation is also evident from the FE contour plot shown in Figure 9. Peak temperature, time to peak temperature, and end-of-cure temperature are quantified as summarized in Table 4. In Table 4, node 5 reported a peak temperature of 113.7°C while node 2 reported the lowest peak temperature of 109.1°C. Such a temperature variation was attributed to the repair patch geometry, where the cure reaction was initiated at the center of the patch, close to node 5. Thus, the center experienced maximum peak exothermic temperatures. This heat was conducted outwards from the center resulting in lower peak temperatures at other nodes. Given the asymmetrical geometry of the patch, node 4 (2:1 slope) presented a distinct temperature profile with a peak temperature of 91.2°C compared to node 6 (1:1 slope), where the peak temperature was 101.5°C. Such variations were also attributed to the geometry of the patch, with node 4 being further away from the center and, consequently, reporting a lower temperature. Nodes 1 and 3 manifested a similar trend as nodes 4 and 6. To further illustrate the influence of the asymmetrical scarfing slope, the temperature of node 9 was also monitored. Note that node 9 was not along the centerline but was 2 mm to the left of the centerline. Node 9 reported the maximum peak temperature of any node within the repair patch, suggesting that the peak exothermic reaction did not initiate along the centerline of the repair but was 2 mm to its left. This was because the uneven scarfing slopes shifted the peak exothermic from the centerline towards the steeper slope suggesting that the repair geometry had a significant influence on the temperature evolution within the repair. Additionally, the time to peak temperature, as seen in Figure 8(a) and Table 4, increased from 66 mins for node 5 to 75 mins for node 1 as the distance of the node increased from the centerline, indicative of a temperature lag in the repair patch. This temperature lag was further evident from the end-of-cure temperature values. Nodes 2, 5, and 7 reported elevated end-of-cure temperatures (> 62°C) compared to nodes 1, 3, 4, and 6 which were much closer to the prescribed cure cycle temperature.

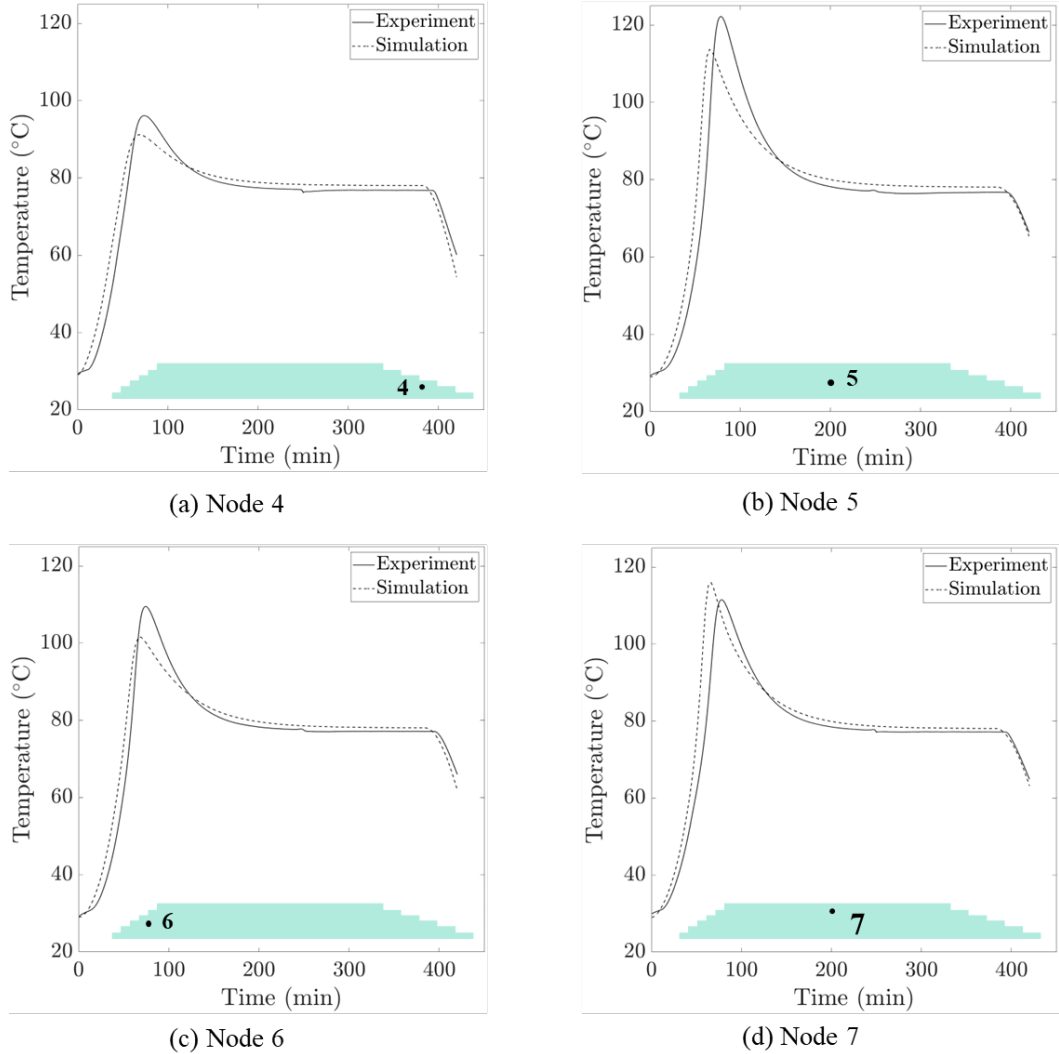

*Figure 6: Plots of temperature evolution in the repair patch as a function of the time for (a) node 4, (b) node 5, (c) node 6, and (d) node 7 from experiments and numerical model.*

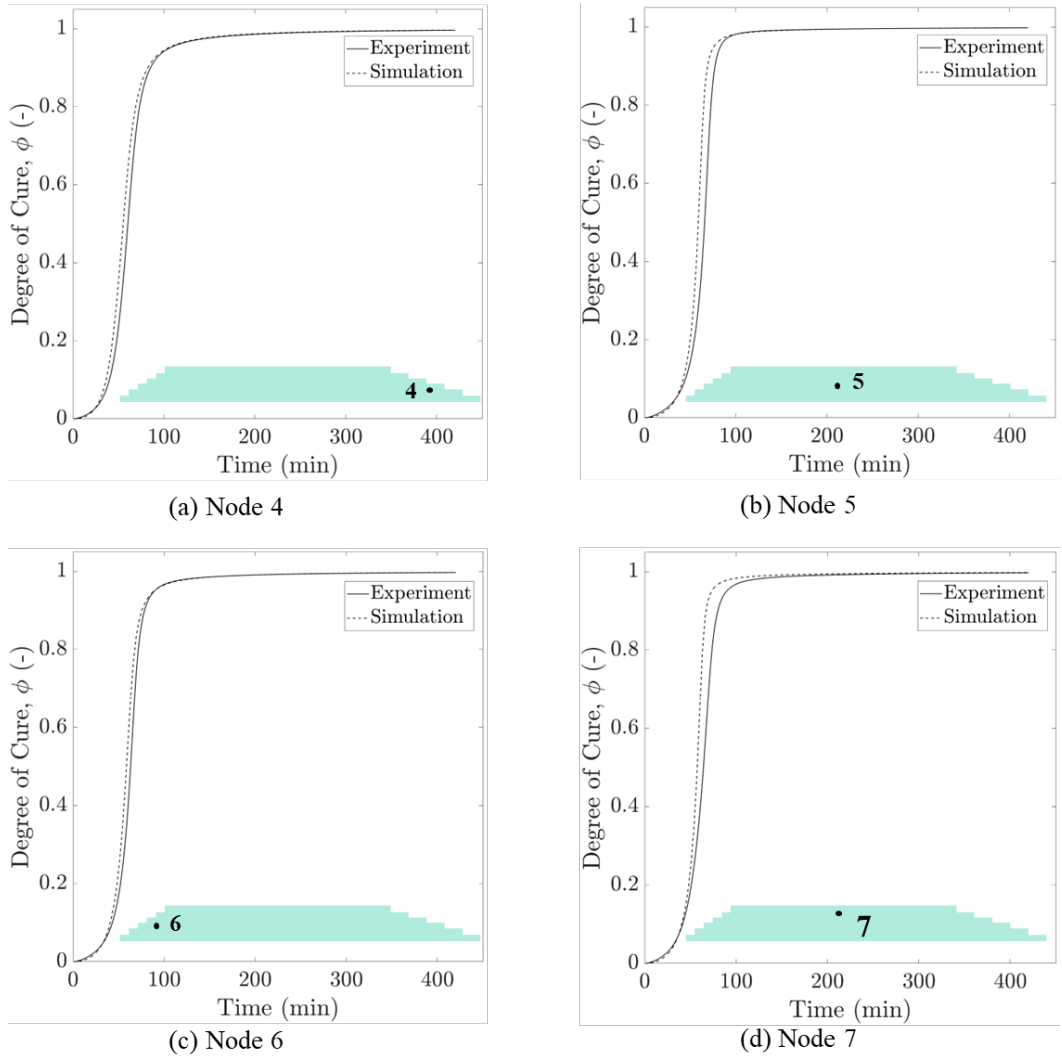

*Figure 7: Plots of the degree of cure evolution in the repair patch as a function of the time for (a) node 4, (b) node 5, (c) node 6, and (d) node 7 from experiments and numerical model.*

The computed degrees of cure for the analyzed nodes are overlaid in Figure 8(b). The degree of cure profiles followed the same trends as the temperature, as expected. Nodes closer to the centerline (nodes 2, 5, 7) were expected to cure faster than the nodes further away from the center (nodes 1, 3, 4, 6) because of their position with respect to the exothermic peak of the reaction. This observation was also evident from Table 4, which summarizes the time taken for each node to reach a complete cure of $\phi = 1$ (indicative of repair time). Node 5 achieved full cure in 153 mins compared to node 2 which reached full cure last after 168 mins. Node 9, which was expected to be closest to the peak exotherm, reached full cure in 144 mins. Nevertheless, all nodes within the laminates reached full cure well before the end of the recommended cure cycle.

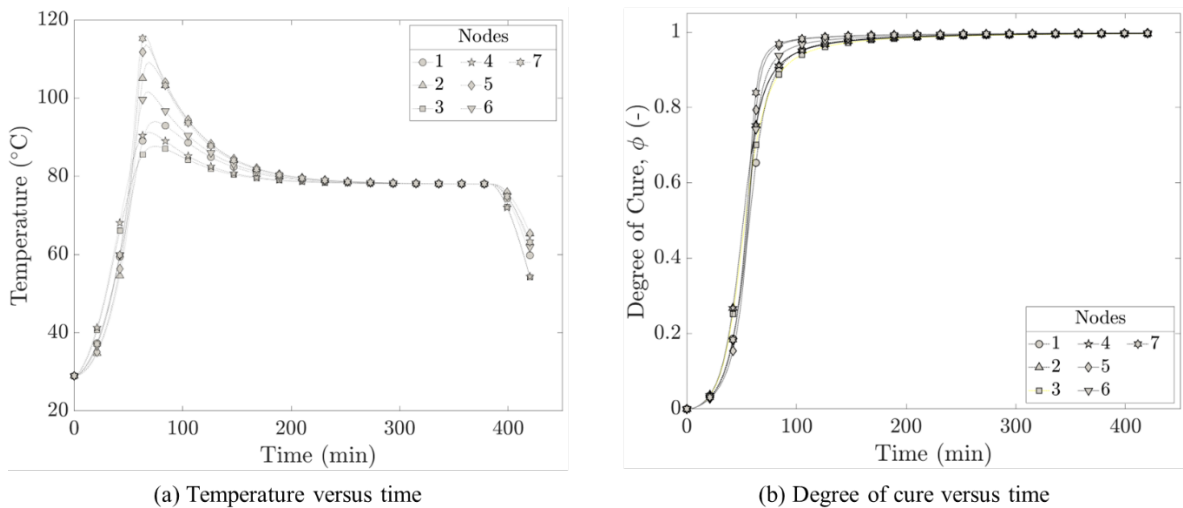

Figure 8: (a) Temperature and (b) degree of cure profiles for all nodes in blade repair simulation.

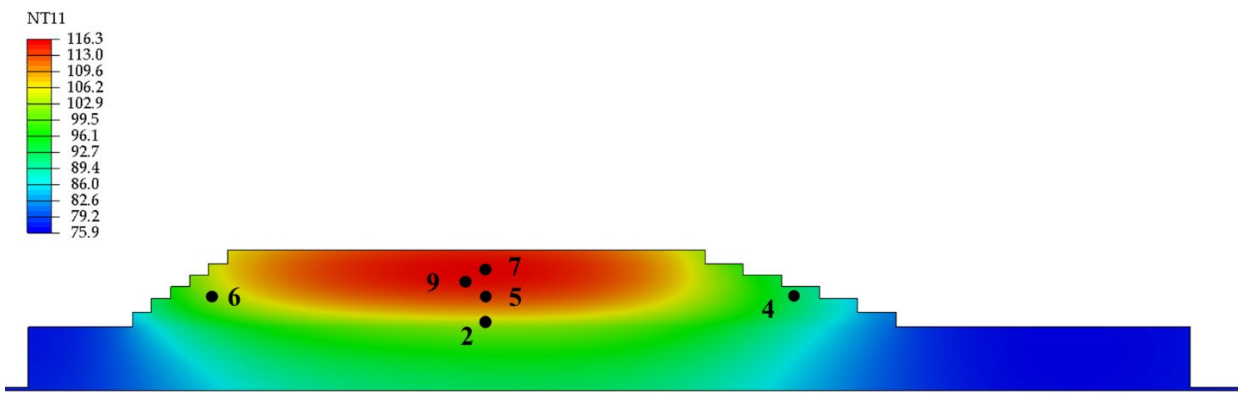

Figure 9: Nodal temperature contour plot of repair patch showing the thermocouple nodes 2, 4, 5, 6, and 7, as well as node 9 experiencing the maximum exotherm at 66 minutes into curing.

*Table 4: Numerical (experimental) peak exothermic temperature, time to peak exothermic temperature, the temperature at the end-of-cure, and time to degree of cure 0.99 for thermocouple nodes 2, 4, 5,6,7, and 9 when no moisture was considered.*

| Nodes | Peak Exotherm (°C) | Time to peak exotherm (min) | End-of-cure temperature (°C) | Time to $\phi = 0.99$ (mins) |
|---|---|---|---|---|
| 2 | 109.1 (113.1) | 69 (85) | 65.4 (67.45) | 168 (224) |
| 4 | 91.2 (96.2) | 69 (73.7) | 54.4 (60.2) | 237 (244) |
| 5 | 113.7 (122.2) | 66 (78.3) | 65.4 (66.4) | 153 (132) |
| 6 | 101.5 (109.5) | 69 (74.3) | 61.9 (66) | 198 (192) |
| 7 | 116.1 (111.6) | 66 (77.3) | 63.2 (64.9) | 147 (180) |
| 9 | 116.3 (-) | 66 (-) | 64.1 (-) | 144 (-) |

### *3.2 Effect of Moisture on Blade Repair*

In the previous sections, the process modeling of a repair scarf laminate was validated against experimental results, and the influence of the repair geometry was evaluated. The effect of moisture content in the resin on temperature and degree of cure evolutions within the patch were evaluated using the same geometry and process modeling technique detailed in Section 2. Various moisture levels, summarized in Table 1, were considered. The corresponding kinetic modeling parameters from Table 2 were used to predict the temperature evolution and degree of cure in the repair patch for different moisture contents. An overlay plot of temperature evolution at several nodes of interest is presented in Figure 10 as a function of the minimum and maximum moisture contents. The influence of moisture on the temperature evolution at nodes 4,5,6, and 7 can be observed in Figure 10; results are also summarized in Table 5 in terms of peak temperature, time to peak temperature, and end-of-cure temperature. For instance, the temperature profile at node 5 for several moisture levels is presented in Figure 11(a). Here the exothermic peak shifted to the left from 66 mins for no moisture case to 42 mins for 12% moisture case indicating an acceleration in the cure reaction with the moisture content. As previously mentioned, adding moisture increases molecular diffusion and dramatically reduces the reaction's activation energy, leading to an early onset of the cure reaction. Concurrently, the magnitude of the peak exothermic temperature dropped with increasing moisture content. Node 5 reported a peak temperature of 113.7°C when no moisture was considered.

However, this value dropped to 88.8°C for the 12% moisture case. This drop in the peak temperature was attributed to the reduced total heat of the reaction with increasing moisture content reported in Table 2. This trend of accelerated cure and reduced peak temperatures with increasing moisture content was consistent for all the nodes considered in Figure 10 (see Table 5). No significant influence of moisture was observed on the end-of-cure nodal temperature.

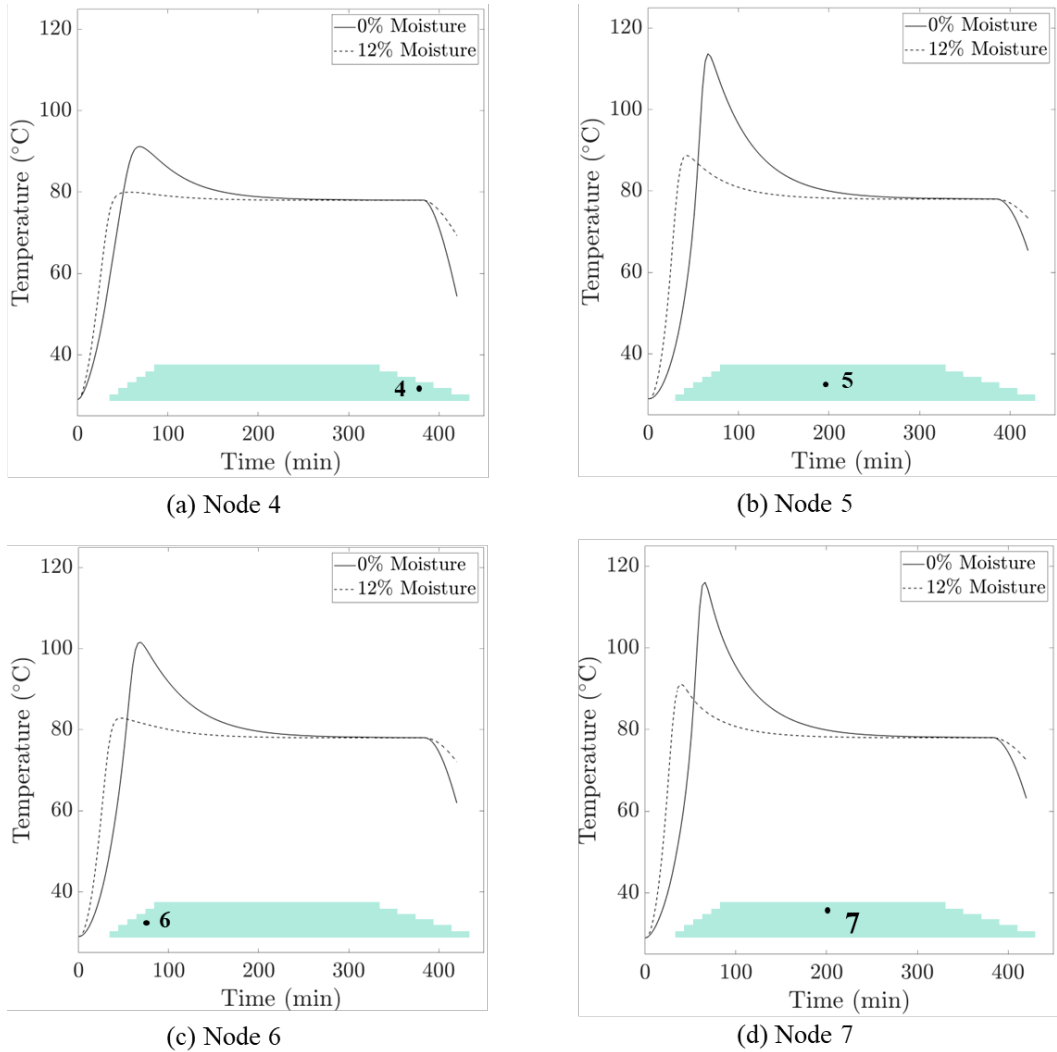

*Figure 10: Plots of temperature evolution in the repair patch as a function of the time and moisture content for (a) node 4, (b) node 5, (c) node 6, and (d) node 7.*

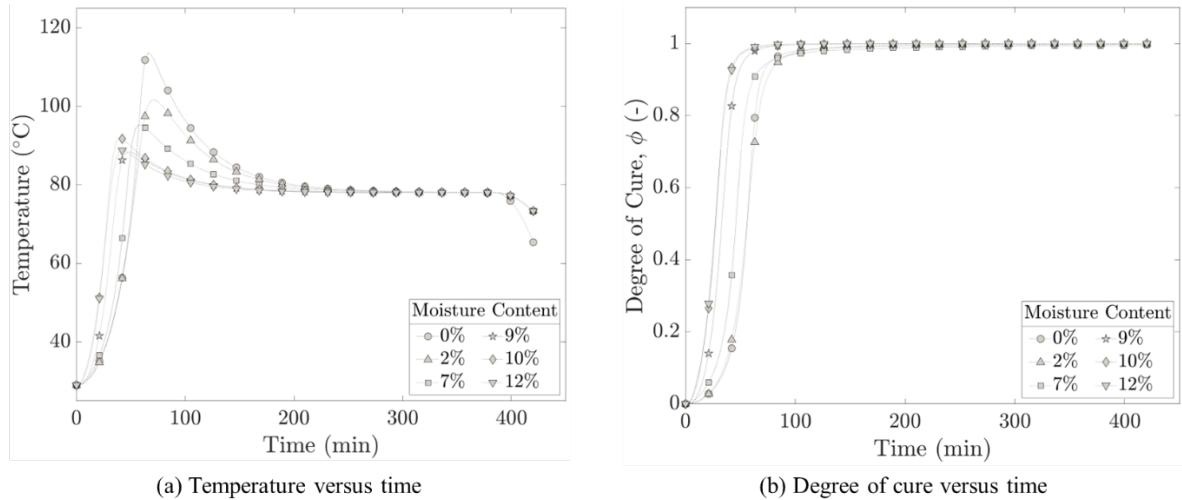

*Figure 11: (a) Temperature and (b) degree of cure profiles for node 5 for six different moisture absorption cases.*

The computed degree of cure for all the nodes of interest as a function of moisture content is presented in Figure 12. Results are also summarized in Table 5. The degree of cure plots exhibited similar and consistent trends to the temperature plots, where the cure rate significantly increased with the increased moisture content. Looking at node 5, for example, the degree of cure started approaching exponential growth after 39 mins of curing when no moisture was considered. By contrast, the 12% moisture case reported the beginning of an exponential cure rate after 12 mins. The faster cure reaction is attributed to enhanced molecular diffusion due to the added moisture [26]. However, after approaching a cure level of approximately 0.89, the cure reaction plateaued for all moisture levels due to the formation of moisture pockets within the free volumes of the crosslinked epoxy [26]. At that point, moisture pockets hindered further molecular diffusion and significantly reduced the reaction rate. However, the epoxy continued to cure until it reached a complete cure. Node 5 achieved full cure after 153 mins when no moisture was added. By contrast, the 12% moisture case took only 63 mins to cure fully. Similar values of time to fully cure were reported for other nodes of interest, as summarized in Table 5.

*Table 5: Peak exothermic temperature, time to peak exothermic temperature, the end-of-cure temperature, and time to degree of cure 0.99 for thermocouple nodes 1-7 and node 9 for all cases of moisture contents.*

| Moisture content | Nodes | Peak Exotherm (°C) | Time to peak exotherm (min) | End-of-cure temperature (°C) | Time to $\phi = 0.99$ (mins) |
|---|---|---|---|---|---|
| 0 % | 2 | 109.1 (113.1) | 69 (85) | 65.4 (67.45) | 168 (224) |
|  | 4 | 91.2 (96.2) | 69 (73.7) | 54.4 (60.2) | 237 (244) |
|  | 5 | 113.7 (122.2) | 66 (78.3) | 65.4 (66.4) | 153 (132) |
|  | 6 | 101.5 (109.5) | 69 (74.3) | 61.9 (66) | 198 (192) |
|  | 7 | 116.1 (111.6) | 66 (77.3) | 63.2 (64.9) | 147 (180) |
|  | 9 | 116.3 (-) | 66 (-) | 64.1 (-) | 144 (-) |
| 2 % | 2 | 99.1 | 75 | 73.4 | 147 |
|  | 4 | 87.0 | 72 | 69.3 | 174 |
|  | 5 | 101.7 | 72 | 73.4 | 141 |
|  | 6 | 93.7 | 72 | 72.1 | 159 |
|  | 7 | 103.0 | 69 | 72.6 | 138 |
|  | 9 | 103.2 | 69.0 | 72.9 | 138 |
| 7 % | 2 | 92.7 | 63 | 73.4 | 240 |
|  | 4 | 83.5 | 66 | 69.3 | 276 |
|  | 5 | 95.3 | 57 | 73.4 | 228 |
|  | 6 | 88.5 | 60 | 72.1 | 258 |
|  | 7 | 97.2 | 57 | 72.6 | 225 |
|  | 9 | 97.2 | 57 | 72.9 | 225 |
| 9 % | 2 | 86.5 | 54 | 73.3 | 78 |
|  | 4 | 80.3 | 60 | 69.3 | 84 |
|  | 5 | 88.4 | 48 | 73.3 | 75 |
|  | 6 | 83.2 | 51 | 72.1 | 81 |
|  | 7 | 90.5 | 45 | 72.5 | 72 |
|  | 9 | 90.3 | 45 | 72.9 | 72 |
| 10 % | 2 | 89.0 | 45 | 73.4 | 75 |
|  | 4 | 81.2 | 51 | 69.3 | 87 |
|  | 5 | 91.9 | 39 | 73.3 | 69 |
|  | 6 | 85.2 | 45 | 72.1 | 81 |
|  | 7 | 94.7 | 39 | 72.5 | 66 |
|  | 9 | 94.5 | 39 | 72.9 | 66 |
| 12 % | 2 | 86.6 | 48 | 73.3 | 66 |
|  | 4 | 80.0 | 57 | 69.3 | 72 |
|  | 5 | 88.8 | 42 | 73.3 | 63 |
|  | 6 | 82.9 | 48 | 72.1 | 69 |
|  | 7 | 91.2 | 39 | 72.5 | 60 |
|  | 9 | 90.9 | 39 | 72.9 | 60 |

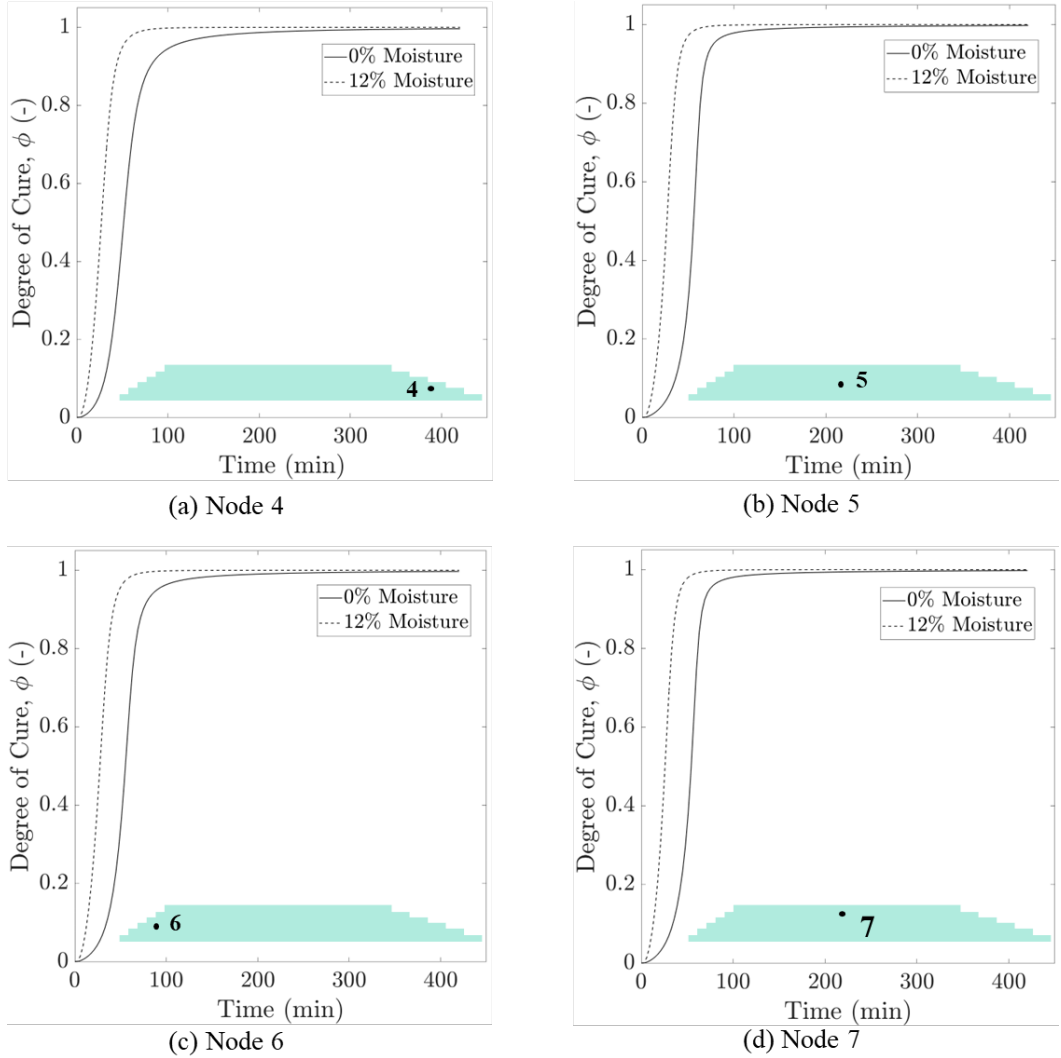

*Figure 12: Plots of the degree of cure evolution in the repair patch as a function of the time and moisture content for (a) node 4, (b) node 5, (c) node 6, and (d) node 7.*

The results presented in this work show that epoxies can absorb a substantial amount of moisture during repairs, which can significantly alter their curing kinetics and affect the evolution of temperature and degree of cure within the repair patch. Furthermore, it was found that the scarfing angle also influenced the spatial variation of the temperature and cure evolution during the repair, with steeper slopes curing faster. Numerical models, like the one presented in this study, can provide an in-depth understanding of the influence of such geometrical and environmental variables on the quality of the repair. Processing conditions, such as heating rates, hold temperatures, and multi-step cure cycles, can be varied, and their influence on the spatial variation of temperature and cure can be analyzed. Considering several parameters, an optimized cure cycle specific to a repair can be determined through the numerical approach presented in this study to ensure even curing in the shortest time.

## 4. Conclusions

This study investigated the influence of environmental moisture on the curing of wind blades during structural repairs. First, the moisture absorption in repair conditions was quantified by individually exposing the polymer and hardener to moisture. It was shown that the curing agent of the two-part resin system absorbed up to 12% moisture by weight when exposed to high temperature and relative humidity. Then, the effect of the moisture on cure kinetics was established by DSC tests. Tests showed that moisture intake significantly accelerated the cure reaction. The effect of altered cure kinetics and repair geometry on the repair time was investigated numerically through FE-based process modeling. The numerical model predictions were validated against a small-scale repair manufactured in the lab in terms of the temperature, degree of cure, and time to complete the repair. It was clear from this study that environmental factors, the size and type of repair involved, and the resin system utilized all play a crucial role in the cure and temperature evolution within the repair, thus making no two repairs alike.

Due to the large number of variables involved, there is a critical need for qualitative and quantitative predictive tools for evaluating optimized cure cycles for each repair. Numerical models, like the one presented in this study, can be used to optimize cure cycles to ensure a rapid, uniform degree of cure throughout the repair. This experimentally validated computational approach will be integrated within an ICME framework for curing-induced residual stress prediction, ultimately enabling a complete connection between materials, reaction kinetics, and mechanical performance.

**Acknowledgments**

This paper is based upon work partially supported by the National Science Foundation under grant numbers 1362022, 1362023, 1916715, and 1916776 (i/UCRC for Wind Energy, Science, Technology, and Research) and from the members of WindSTAR I/UCRC. Any opinions, findings, and conclusions or recommendations expressed in this material are those of the author(s) and do not necessarily reflect the view of the Nation Science Foundations or the sponsors. The authors would also like to thank the WindSTAR Industrial Advisory Board members Steve Nolet and Amir Salimi (TPI Composites); Nathan Bruno, Mirna Robles, and Paul Ubrich (Westlake Epoxy); Ben Rice (Pattern Energy); and Jian Lahir (EDPR) for their technical advice and support and for providing the resin for experimental investigation.

The authors declare no conflict of interest.

# References

[1] K. B. Katnam, A. J. Comer, D. Roy, L. F. M. da Silva, and T. M. Young, "Composite Repair in Wind Turbine Blades: An Overview," *The Journal of Adhesion*, vol. 91, no. 1–2, pp. 113–139, Jan. 2015, doi: 10.1080/00218464.2014.900449.

[2] L. Mishnaevsky, K. Branner, H. Petersen, J. Beauson, M. McGugan, and B. Sørensen, "Materials for Wind Turbine Blades: An Overview," *Materials*, vol. 10, no. 11, p. 1285, Nov. 2017, doi: 10.3390/ma10111285.

[3] L. Mishnaevsky, N. Frost-Jensen Johansen, A. Fraisse, S. Fæster, T. Jensen, and B. Bendixen, "Technologies of Wind Turbine Blade Repair: Practical Comparison," *Energies*, vol. 15, no. 5, Art. no. 5, Jan. 2022, doi: 10.3390/en15051767.

[4] M. Li, "TEMPERATURE AND MOISTURE EFFECTS ON COMPOSITE MATERIALS FOR WIND TURBINE BLADES," Montana State University-Bozeman, Bozeman Montana, 2000.

[5] A. Candela Garolera, S. F. Madsen, M. Nissim, J. D. Myers, and J. Holboell, "Lightning Damage to Wind Turbine Blades From Wind Farms in the U.S.," *IEEE Trans. Power Delivery*, vol. 31, no. 3, pp. 1043–1049, Jun. 2016, doi: 10.1109/TPWRD.2014.2370682.

[6] M. N. Olaya, J. Mcdonald, S. Shah, C. J. Hansen, S. E. Stapleton, and M. Maiaru, "Wind Blade Repair Optimization," in *Proceedings of the American Society for Composites — Thirty-fifth Technical Conference*, Sep. 2020. doi: 10.12783/asc35/34979.

[7] L. Mischnaewski and L. Mishnaevsky, "Structural repair of wind turbine blades: Computational model for the evaluation of the effect of adhesive properties," *Wind Energy*, vol. 24, no. 4, pp. 402–408, Apr. 2021, doi: 10.1002/we.2575.

[8] L. Mishnaevsky, "Repair of wind turbine blades: Review of methods and related computational mechanics problems," *Renewable Energy*, vol. 140, pp. 828–839, Sep. 2019, doi: 10.1016/j.renene.2019.03.113.

[9] S. P. Shah, S. U. Patil, C. J. Hansen, G. M. Odegard, and M. Maiarù, "Process modeling and characterization of thermoset composites for residual stress prediction," *MAMS*, pp. 1–12, Dec. 2021, doi: 10.1080/15376494.2021.2017527.

[10] S. Anandan, G. S. Dhaliwal, Z. Huo, K. Chandrashekhara, N. Apetre, and N. Iyyer, "Curing of Thick Thermoset Composite Laminates: Multiphysics Modeling and Experiments," *Appl Compos Mater*, vol. 25, no. 5, pp. 1155–1168, Oct. 2018, doi: 10.1007/s10443-017-9658-9.

[11] K. Bujun, "Processing Study of in-situ Bonded Scarf Repairs for Composite Structures," McGIll University, Montreal, Quebec, 2004.

[12] A. G. Cassano, S. Dev, M. Maiaru, C. J. Hansen, and S. E. Stapleton, "Cure simulations of thick adhesive bondlines for wind energy applications," *Journal of Applied Polymer Science*, vol. 138, no. 10, p. 49989, 2021, doi: 10.1002/app.49989.

[13] C. Heinrich, "The Influence of the Curing Process on the Response of Textile Composites," University of Michigan, Ann Arbor, 2011.

[14] X. Li, J. Wang, S. Li, and A. Ding, "Cure-induced temperature gradient in laminated composite plate: Numerical simulation and experimental measurement," *Composite Structures*, vol. 253, p. 112822, Dec. 2020, doi: 10.1016/j.compstruct.2020.112822.

[15] A. Loos and G. Springer, "Calculation of Cure Process Variables During Cure of Graphite/Epoxy Composites," in *Composite Materials: Quality Assurance and Processing*, C. Browning, Ed., 100

Barr Harbor Drive, PO Box C700, West Conshohocken, PA 19428-2959: ASTM International, 1983, pp. 110-110–9. doi: 10.1520/STP28542S.

[16] S. Nakouzi, J. Pancrace, F. M. Schmidt, Y. Le Maoult, and F. Berthet, "Curing Simulation of Composites Coupled with Infrared Heating," *Int J Mater Form*, vol. 3, no. S1, pp. 587–590, Apr. 2010, doi: 10.1007/s12289-010-0838-5.

[17] F. H. Darwish and K. N. Shivakumar, "Experimental and Analytical Modeling of Scarf Repaired Composite Panels," *Mechanics of Advanced Materials and Structures*, vol. 21, no. 3, pp. 207–212, Mar. 2014, doi: 10.1080/15376494.2013.834096.

[18] C. H. Wang, V. Venugopal, and L. Peng, "Stepped Flush Repairs for Primary Composite Structures," *The Journal of Adhesion*, vol. 91, no. 1–2, pp. 95–112, Jan. 2015, doi: 10.1080/00218464.2014.896212.

[19] C. H. Wang and A. J. Gunnion, "On the design methodology of scarf repairs to composite laminates," *Composites Science and Technology*, vol. 68, no. 1, pp. 35–46, Jan. 2008, doi: 10.1016/j.compscitech.2007.05.045.

[20] C. Soutis and F. Z. Hu, "Failure Analysis of Scarf-Patch-Repaired Carbon Fiber/Epoxy Laminates Under Compression," *AIAA Journal*, vol. 38, no. 4, pp. 737–740, Apr. 2000, doi: 10.2514/2.1027.

[21] S. P. Shah and M. Maiarù, "Effect of Manufacturing on the Transverse Response of Polymer Matrix Composites," *Polymers*, vol. 13, no. 15, Art. no. 15, Jan. 2021, doi: 10.3390/polym13152491.

[22] M. Maiaru, "Effect of uncertainty in matrix fracture properties on the transverse strength of fiber reinforced polymer matrix composites," in *2018 AIAA/ASCE/AHS/ASC Structures, Structural Dynamics, and Materials Conference*, American Institute of Aeronautics and Astronautics, 2018. doi: 10.2514/6.2018-1901.

[23] M. Maiarù, R. J. D'Mello, and A. M. Waas, "Characterization of intralaminar strengths of virtually cured polymer matrix composites," *Composites Part B: Engineering*, vol. 149, pp. 285–295, Sep. 2018, doi: 10.1016/j.compositesb.2018.02.018.

[24] L. Aktas, Y. Hamidi, and M. C. Altan, "Effect of Moisture Absorption on Mechanical Properties of Resin Transfer Molded Composites," in *Materials: Processing, Characterization and Modeling of Novel Nano-Engineered and Surface Engineered Materials*, New Orleans, Louisiana, USA: ASMEDC, Jan. 2002, pp. 173–181. doi: 10.1115/IMECE2002-39223.

[25] S. Choi, A. P. Janisse, C. Liu, and E. P. Douglas, "Effect of water addition on the cure kinetics of an epoxy-amine thermoset," *Journal of Polymer Science Part A: Polymer Chemistry*, vol. 49, no. 21, pp. 4650–4659, 2011, doi: 10.1002/pola.24909.

[26] N. Sharp, C. Li, A. Strachan, D. Adams, and R. B. Pipes, "Effects of water on epoxy cure kinetics and glass transition temperature utilizing molecular dynamics simulations," *Journal of Polymer Science Part B: Polymer Physics*, vol. 55, no. 15, pp. 1150–1159, 2017, doi: 10.1002/polb.24357.

[27] C.-H. Shen and G. S. Springer, "Moisture Absorption and Desorption of Composite Materials," *Journal of Composite Materials*, vol. 10, 1976.

[28] J. M. Kropka, D. B. Adolf, S. Spangler, K. Austin, and R. S. Chambers, "Mechanisms of degradation in adhesive joint strength: Glassy polymer thermoset bond in a humid environment," *International Journal of Adhesion and Adhesives*, vol. 63, pp. 14–25, Dec. 2015, doi: 10.1016/j.ijadhadh.2015.07.014.

[29] S. Patil *et al.*, "Multi-scale Approach to Predict Cure-Induced Residual Stresses in an Epoxy System," in *Proceedings of the American Society of Composites 35th Technical (Virtual) Conference*, American Society of Composites, Sep. 2020.

[30] P. P. Deshpande *et al.*, "Multiscale Modelling of the Cure Process in Thermoset Polymers Using ICME," in *Proceedings of the American Society for Composites " Thirty-fifth Technical Conference*, 2020. doi: 10.12783/asc35/34889.

[31] G. M. Odegard *et al.*, "Molecular Dynamics Modeling of Epoxy Resins Using the Reactive Interface Force Field," *Macromolecules*, vol. 54, no. 21, pp. 9815–9824, Nov. 2021, doi: 10.1021/acs.macromol.1c01813.

[32] S. Shah *et al.*, "Multiscale Modeling for Virtual Manufacturing of Thermoset Composites," in *AIAA Scitech 2020 Forum*, in AIAA SciTech Forum. American Institute of Aeronautics and Astronautics, Jan. 2020. doi: 10.2514/6.2020-0882.

[33] S. U. Patil, S. P. Shah, M. Olaya, P. P. Deshpande, M. Maiaru, and G. M. Odegard, "Reactive Molecular Dynamics Simulation of Epoxy for the Full Cross-Linking Process," *ACS Appl. Polym. Mater.*, Oct. 2021, doi: 10.1021/acsapm.1c01024.

[34] P. S. Gaikwad *et al.*, "Understanding the Origin of the Low Cure Shrinkage of Polybenzoxazine Resin by Computational Simulation," *ACS Appl. Polym. Mater.*, vol. 3, no. 12, pp. 6407–6415, Dec. 2021, doi: 10.1021/acsapm.1c01164.

[35] R. J. D'Mello, M. Maiarù, and A. M. Waas, "Effect of the curing process on the transverse tensile strength of fiber-reinforced polymer matrix lamina using micromechanics computations," *Integrating Materials*, vol. 4, no. 1, pp. 119–136, Dec. 2015, doi: 10.1186/s40192-015-0035-y.

[36] S. Shah and M. Maiaru, "Microscale Analysis of Virtually Cured Polymer Matrix Composites Accounting for Uncertainty in Matrix Properties During Manufacturing," in *American Society for Composites 2018*, DEStech Publications, Inc., Nov. 2018. doi: 10.12783/asc33/25958.

[37] R. J. D'Mello, A. M. Waas, M. Maiaru, and R. Koon, "Integrated Computational Modeling for Efficient Material and Process Design for Composite Aerospace Structures," in *AIAA Scitech 2020 Forum*, American Institute of Aeronautics and Astronautics. doi: 10.2514/6.2020-0655.

[38] M. N. Olaya and M. Maiaru, "A multi-scale approach for process modeling of polymer matrix composites," in *AIAA SCITECH 2022 Forum*, American Institute of Aeronautics and Astronautics. doi: 10.2514/6.2022-0379.